\documentclass[doublecol]{epl2}
\usepackage{graphicx}
\usepackage{amsmath}

\title{Phase-Tunable temperature amplifier}

\author{F. Paolucci\inst{1} \and G. Marchegiani\inst{1,2} \and E. Strambini\inst{1} \and F. Giazotto\inst{1}}
\shortauthor{F. Author \etal}

\institute{                    
  \inst{1} NEST, Instituto Nanoscienze-CNR and Scuola Normale Superiore, I-56127 Pisa, Italy\\
  \inst{2} Dipartimento di Fisica dell'Universit\`a di Pisa, Largo Pontecorvo 3, I-56127 Pisa, Italy
}

\abstract{Coherent caloritronics, the thermal counterpart of coherent electronics, has drawn growing attention since the discovery of heat interference in $2012$. Thermal interferometers, diodes, transistors and nano-valves have been theoretically proposed and experimentally demonstrated by exploiting the quantum phase difference between two superconductors coupled through a Josephson junction. So far, the quantum-phase modulator has been realized in the form of a superconducting quantum interference device (SQUID) or a superconducting quantum interference proximity transistor (SQUIPT). Thence, an external magnetic field is necessary in order to manipulate the heat transport. Here, we theoretically propose the first on-chip fully thermal caloritronic device: the phase-tunable temperature amplifier (PTA). Taking advantage of a recently discovered thermoelectric effect in spin-split superconductors coupled to a spin-polarized system, we generate the magnetic flux controlling the transport through a temperature biased SQUIPT by applying a temperature gradient. We simulate the behavior of the device and define a number of figures of merit in full analogy with voltage amplifiers. Notably, our architecture ensures almost infinite input thermal impedance, maximum gain of about $11$ and efficiency reaching the $95\%$. This concept paves the way for applications in radiation sensing, thermal logics and quantum information.}

\begin{document}

\maketitle

\section{Introduction}
The discovery of thermoionic emission by Fredrick Guthrie in 1873 \cite{Guthrie1876} brought to the invention of the first electronic devices: the diode and triode amplifiers \cite{Guarnieri2012}. After more than $100$ years, the recent advances of transistor-based technology \cite{Pugh1991} made possible the design and production of new daily life devices. In the era of energy saving, the common goal in electronics is to increase the device efficiency in order to abate energy losses and pollutant emissions. Anyways, further developments of nowadays technology are bounded by quantum mechanical restrictions to miniaturization and by heat dissipation \cite{Mannhart2010}. The inescapable heat generated in solid-state nano-structures is considered detrimental in electronics. As a consequence, the ability of mastering the heat transport in such structures has been only recently investigated \cite{Giazotto2006}, and it could lead to new concepts and capabilities. In this framework, the experimental demonstration in $2012$ of heat interference in a SQUID \cite{Giazotto2012} heralded the foundation of the thermal counterpart of coherent electronics: coherent caloritronics \cite{Martinez2014, Mart2014}. Despite it is still distant from the ripeness of electronics, coherent caloritronics is rapidly growing through the design and the realization of thermal analogues of electronic devices, such as heat diodes \cite{Giazotto2012}, transistors \cite{Fornieri2015}, valves \cite{Strambini2014}, amplifiers \cite{Fornieri2016} and modulators \cite{Giaz2012}. One of the theoretical foundations of coherent caloritronics resides in the prediction of the periodic dependence of thermal currents across a Josephson junction \cite{Josephson1962} on the quantum phase difference between the two superconductors \cite{Maki1965}. Hence, the resulting thermal modulation acquires a phase-coherent character. So far, quantum interference between Josephson-coupled superconductors has been realized through the use of a SQUID \cite{Clarke} or, more recently, taking advantage from a newly designed SQUIPT \cite{Giazotto2010, Meschke2011, Virtanen2016}. Thereby, the thermal transport across caloritronic devices is manipulated by a magnetic flux $\Phi$ threading a superconducting ring, and an external source of magnetic field is essential. The last requirement impeded the realization of fully thermal on-chip coherent caloritronic devices up to now. In the last two years, surprisingly large thermoelectric effects in spin-filtered superconducting tunnel junctions have been predicted \cite{Machon2013, Ozaeta2014} and demonstrated \cite{Kolenda2016}. This discovery enables the direct transduction, for the first time at cryogenic temperatures, of temperature gradients into electrical signals. 
\begin{figure*}[ht]
  \includegraphics {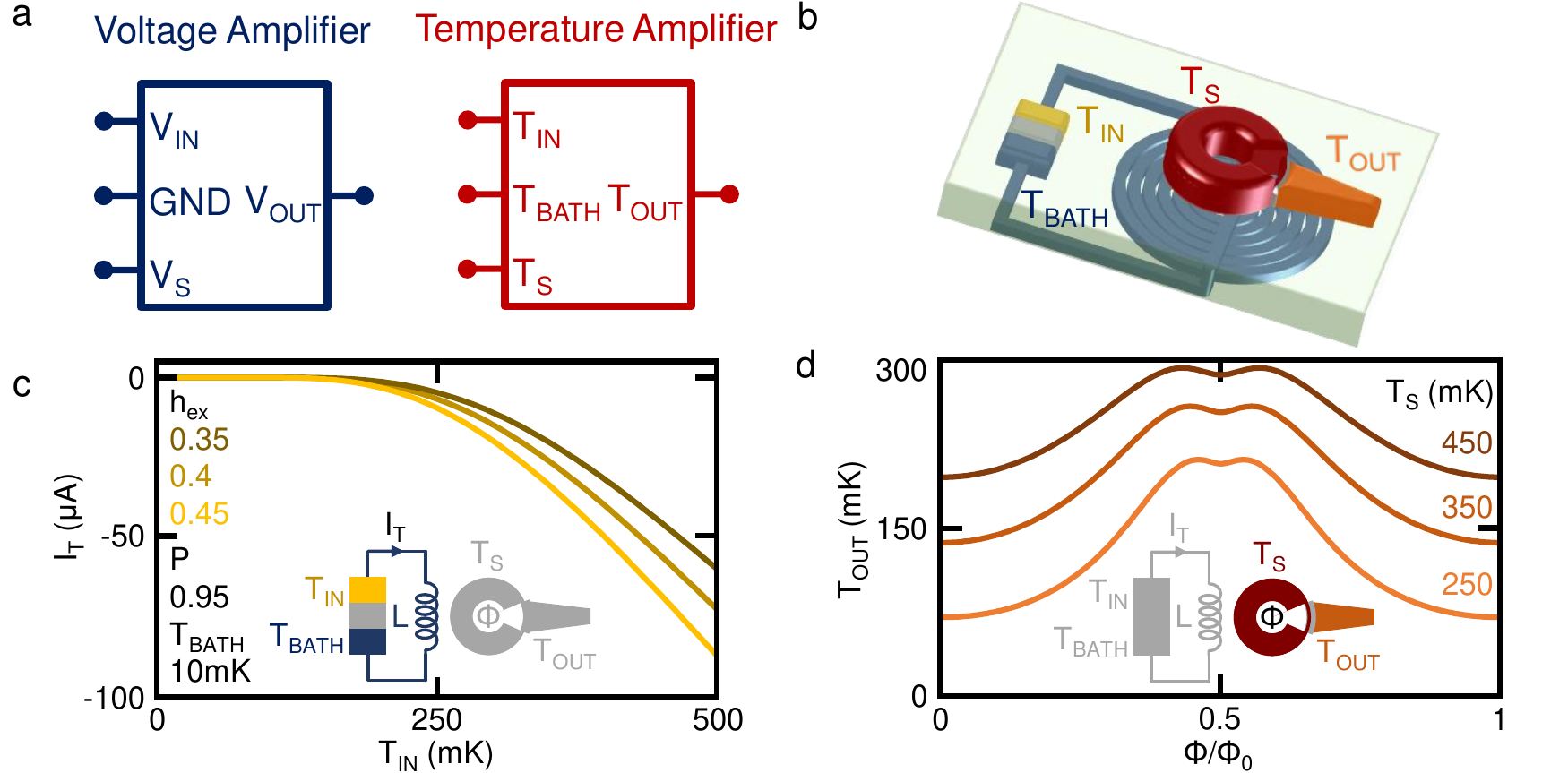}
  \caption{(a) Circuit diagram symbols of voltage amplifier (blue) and temperature amplifier (red). The input ($V_{IN}$ and $T_{IN}$), reference ($V_{REF}$ and $T_{BATH}$) and output ($V_{OUT}$ and $T_{OUT}$) signals, and the power supplies ($V_{S}$ and $T_{S}$) are represented. (b) Schematic representation of the temperature amplifier: the thermoelectric element highlighted with the dashed rectangle is constituted of a metal (yellow), a ferromagnetic insulator (gray) and a superconductor (turquoise). The turquoise depicts the superconducting coil. The SQUIPT is composed of a $S_1N_1$ ring (red) and a tunnel-coupled metal probe (orange) through a thin insulator (dark gray). (c) Closed circuit thermoelectric current $I_T$ as a function of $T_{IN}$ for different values of $h_{ex}$.(d) Output temperature $T_{OUT}$ of the SQUIPT as a function of $\Phi$ for different values of $T_S$. Parameters are listed in the Appendices.}
  \label{Figure1}
\end{figure*}
Here we present the first on-chip fully thermal device in caloritronics: the phase-tunable temperature amplifier (PTA). 
Our architecture takes advantage from the closed-circuit current generated by a thermoelectric element in order to create a magnetic field which controls heat transport across a thermal nano-valve. By employing widely used materials and a geometry feasible with standard lithographic techniques, we show the basic input-to-output temperature conversion, and define several figures of merit in analogy to electronics to evaluate the performances of the temperature amplifier. The device layout may foster its use in different field of science, like quantum information \cite{Nielsen}, thermal logics \cite{Li2012} and radiation detection \cite{Giazotto2006}.
\section{Working principle and basic behavior}
The PTA is the caloritronic equivalent of the voltage amplifier in electronics \cite{Millman}, since temperature is the thermal counterpart of electric potential. The voltage-temperature analogy is schematized in fig. \ref{Figure1}-a, where the usual symbol of voltage amplifiers (blue) and the corresponding representation of temperature amplifiers (red) are depicted. A voltage amplifier is a device which produces an output signal $V_{OUT}=G~\Delta V_{IN}$ , where $G>1$ is the gain and $\Delta V_{IN}=V_{IN}-V_{REF}$ is the difference between the input signal $V_{IN}$ and the reference $V_{REF}$. Since the law of conservation of energy does not allow the creation of energy, the system requires a voltage supply $V_S$ to operate. Analogously, a temperature amplifier generates an output temperature $T_{OUT}=G~T_{IN}$, where $T_{IN}$ is the input signal. In this case, the operation power is supplied by a temperature $T_S$. Differently from electronics, where the absolute value of the signals has no physical meaning and an arbitrary reference potential is required, in caloritronics the temperature signals can take only positive values and they are always referred to zero temperature (zero energy). Thereby, the base temperature $T_{BATH}$ has a different and more complex role than a simple reference. It defines the background energy level, the operation \cite{Strambini2014} and the energy losses of the system due to electron-phonon interaction \cite{Giazotto2006}. In the following, we set $T_{BATH}=10~$mK that ensures  low noise and reduced energy losses.
The PTA is composed of a normal metal-ferromagnetic insulator-superconductor ($N-FI-S$) tunnel junction inductively coupled to a SQUIPT \cite{Giazotto2010} through a superconducting coil.

In an electronic conductor, a thermoelectric effect can be generated by breaking the electron-hole symmetry in the  density of states (DOS) \cite{Mermin}. Recently, it has been shown that this can be efficiently realized in superconductor-based structure: i) by inducing a Zeeman spin-splitting $h_{ex}$ in the quasiparticle DOS, hence breaking the electron-hole symmetry for each spin band, ii) by selecting a specific spin band (spin-filtering) \cite{Giaz2008, Ozaeta2014}. In our scheme, both the mechanisms are provided by a single ferromagnetic insulator layer of the $N-FI-S$ junction \cite{Giazotto2015}. A temperature gradient between the normal metal $N$ and the superconductor $S$ generates the thermoelectric signal: an open circuit thermovoltage $V_T$  in the Seebeck regime or a closed circuit thermocurrent $I_T$ in the Peltier regime \cite{Giazotto2015}. In our device, we take advantage of the closed circuit thermocurrent in order to create a magnetic field by means of a superconducting coil of self-inductance $L$. The superconductor is kept at $T_{BATH}$ while the normal metal is set to the input temperature $T_{IN}>T_{BATH}$, because in this configuration the provided thermocurrent exhibits a monotonic behavior with rising temperature gradient \cite{Giazotto2015}. Figure \ref{Figure1}-c shows the dependence of $I_T$ on $T_{IN}$ for different values of $h_{ex}$. The thermocurrent is a growing function of the spin-splitting of the DOS (i.e. $h_{ex}$) and abruptly rises when the thermal gradient is greater than a critical value (in our numerical calculation $T_{IN}\geq$200 mK). The detailed description of the temperature-to-current transduction of the $N-FI-S$ junction is given in the Appendices.
\begin{figure*}[ht]
  \includegraphics {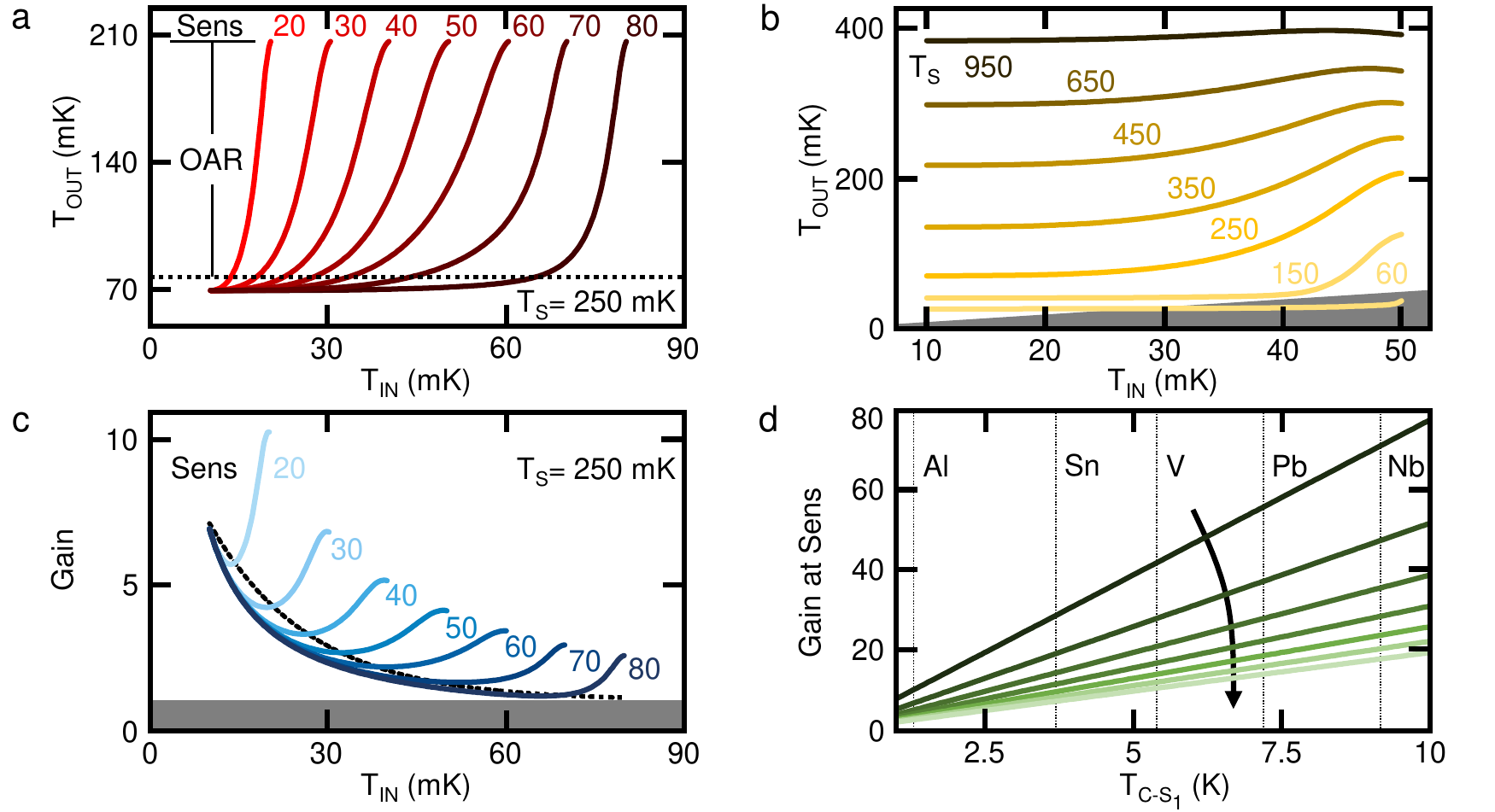}
  \caption{(a) Output temperature $T_{OUT}$ as a function of $T_{IN}$ calculated for  $T_S=250~$mK and for different values of $Sens$. The black dotted line represents the minimum value of active output $T_{OUT_{MIN}}$. The output active range $OAR$ is shown. (b) Output temperature $T_{OUT}$ as a function of $T_{IN}$ calculated for different values of $T_S$. The gray triangle depicts the portion of the parameters space with $G\leq1$. (c) Gain $G$ as a function of $T_{IN}$  calculated for $T_S=250~$mK and for different values of $Sens$. The gray rectangle represents the area of $G\leq1$. The black dotted line represents the minimum value of $OAR$. (d) Gain $G$ as a function of $T_{C-S_1}$ for a constant ratio $T_{C-S_1}/T_S=5.2$ and different values $Sens$ (increasing with the arrow direction). Cuts at critical temperatures of relevant superconducting materials are represented.}
  \label{Figure2}
\end{figure*}

We now turn our attention on the second building block of our device: the thermally biased SQUIPT. It is composed of a normal metal wire $N_1$ interrupting a superconducting ring $S_1$, as portrayed in fig. \ref{Figure1}-b. Owing to the good electric contact between $N_1$ and $N_2$, the metal wire acquires a superconducting character through the superconducting proximity effect \cite{Holm1932}. A normal metal $N_2$ probe tunnel-coupled to the wire through a thin insulating layer acts as the output electrode of the device. A magnetic flux $\Phi$ threading the ring modulates the density of states of the proximized wire \cite{Petrashov1995, leSueur2008} and, as a consequence, the electronic transport between $N_1$ and $N_2$ \cite{Giazotto2010,  Meschke2011}. Analogously, the temperature-biased SQUIPT has been predicted to act as a thermal nano-valve leading to a phase-dependent thermal transport between $S_1$ and $N_1$ \cite{Strambini2014}. The detailed theoretical description of the SQUIPT can be found in the Appendices. The thermal behavior of the nano-valve is resumed in fig. \ref{Figure1}-d, where the dependence of $T_{OUT}$ on the magnetic flux $\Phi$ for different values of $T_S$ is plotted. The probe temperature is minimum at $\Phi=0$, where the energy gap is fully induced in the $N_1$ DOS. When the magnetic field is switched on, the probe temperature increases due to the closure of the minigap \cite{Giazotto2010}, reaching a maximum  at $\Phi\sim0.45~\Phi_0$ and slightly lowering for $\Phi \to \Phi_0/2$ \cite{Strambini2014}. Furthermore, the maximum value of $T_{OUT}$ increases with $T_S$ while its modulation with $\Phi$ softens for large values of the supply temperature. Notably, thermal transport across the SQUIPT is phase-dependent, because it is modulated by the superconducting macroscopic phase difference across the proximized wire \cite{Strambini2014}.

The architecture of the PTA requires to couple these two building blocks. This goal is achieved by means of a superconducting coil of inductance $L$ connected to the thermoelectric element (see fig. \ref{Figure1}-b). By placing the thermal nano-valve in the center of this coil is possible to drive the SQUIPT by means of the static magnetic flux generated by the coil. The magnetic flux through the SQUIPT is $\Phi=M~I_T$, where $M$ is the mutual inductance between the coil and the SQUIPT. This assembly permits to relate the input $T_{IN}$ with the output $T_{OUT}$ temperature. As typically done in electronics, it is useful to introduce a parameter which sets the input corresponding to the maximum operating output required (here $T_{OUT_{MAX}}$). This quantity is tipically called sensitivity (here we use the symbol $Sens$). As already seen (fig 1-d), the temperature of the output probe $N_2$ increases monotonically with the flux for values smaller than $\sim0.45\Phi_0$, where it reaches a maximum. Furthermore, the thermocurrent, hence the flux, increases monotonically with the input temperature. If we define $I_{T_{MAX}}$ as the current generated by the thermoelectric temperature for $T_{IN}=Sens$, the coupling required is $M=0.45\Phi_0/I_{T_{MAX}}$ and the output is a growing function of the input signal, as normally required to an amplifier. Note that the coupling inductance scales inversely with the sensitivity (i.e. $Sens\sim 1/M$). This is not surprising: if we consider a high operating temperature (high $Sens$), a low thermocurrent is sufficient to perform the job.

The basic behavior of the temperature amplifier is illustrated in fig. \ref{Figure2}-a, where the dependence of the output temperature $T_{OUT}$ on the input temperature $T_{IN}$ is depicted for a supply temperature $T_S=250~$mK and for different sensitivities $Sens$. Note that both the minimum and the maximum output temperature are independent on $Sens$. The minimum temperature is obtained at null input signal, i.e. when the normal layer $N$ of the $N-FI-S$ element is at the bath temperature $T_{IN}=T_{BATH}$. For this reason, we refer to it as noise temperature $T_{Noise}$. The maximum, by definition, is obtained at $T_{IN}=Sens$, corresponding to a flux $0.45\Phi_0$. The horizontal dotted black line sets the minimum value of the output active range $OAR=T_{OUT_{MAX}}-T_{{OUT}_{MIN}}$ (i.e. the interval where the output varies with the input signal), defined as $T_{{OUT}_{MIN}}=T_{Noise}+10\%T_{Noise}$. The size of the OAR is independent on $Sens$ (for our simulation parameters is approximately 130 mK). The independence of the OAR on the $Sens$ may appear surprising at first. However it is easy to understand once it is realized that the OAR is only related on the valve (SQUIPT) operation, whereas $Sens$ only affects the coupling required between the thermoelectric and the valve. On the other hand, $T_{OUT}$ calculated at a specific $T_{IN}$ drops by increasing $Sens$, because the $I_T$ is independent of the sensitivity, and the inducting coupling $M$ lowers by increasing $Sens$. 

The supply temperature $T_S$ has a great influence on the behavior of the PTA, because it defines the minimum and the maximum values of $T_{OUT}$, as illustrated in fig. \ref{Figure2}-b. For values of $T_S$ comparable to the critical temperature  $T_{C-S_1}$ of the ring of the SQUIPT, $T_{OUT}$ depends only weakly on $T_{IN}$, because the energy gap of the ring $\Delta_{S_1}$ closes and the proximized wire assumes an almost metallic character for every value of the magnetic flux $\Phi$ (i.e input temperature $T_{IN}$). By lowering $T_S$ the superconducting pairing potential rises and the flux $\Phi$ successfully modulates thermal transport across the SQUIPT in the complete range $0-0.45~\Phi_0$, hence the output temperature varies with all the values of the input signal (see the traces for $T_S=450-150~$mK in fig. \ref{Figure2}-b). When $T_S\leq0.1~ T_{C-S_1}$ the thermal broadening of the Fermi distribution $k_B T_S$ is small compared to the energy gap of the ring, and the phase dependence of the thermal transport becomes dominant only when the energy gap is almost fully suppressed, i.e. $\Phi \to 0.45~\Phi_0$. Thereby, the output temperature is exclusively modulated for $T_{IN} \approx Sens$ and the output signal can be lower than the input, as shown for $T_S=60~$mK in fig. \ref{Figure2}-b. The ensemble of these behaviors leads to the conclusion that the temperature amplifier efficiently works when $0.1~ T_{C-S_1}\leq T_S \leq 0.4~ T_{C-S_1}$. 

The most relevant parameter for an amplifier is the gain $G$, which is plotted in fig. \ref{Figure2}-c as a function of the $T_{IN}$ for different values of $Sens$ and $T_S=250~$mK. The gain is independent on $Sens$ for $T_{IN}=T_{BATH}$, because $T_{Noise}$ is only determined by $T_S$. On the contrary, $G$ strongly depends on $Sens$ when the output temperature resides in the $OAR$ (i.e. $T_{OUT}\geq T_{OUT_{MIN}}$). In particular, $G$ lowers by increasing sensitivity at fixed $T_{IN}$, and $G(T_{IN}=Sens)$ drops for rising $Sens$, because $M$ scales inversely with the sensitivity and the maximum output signal is exclusively controlled by $T_S$ (see fig. \ref{Figure2}-a). For a given $Sens$, the gain grows with $T_{IN}$ when the amplifier is in the active output mode, i.e. the values of $G$ above the black dotted line in fig. \ref{Figure2}-c. This behavior is the result of the joint action of the temperature-to-current conversion due to the thermoelectric element and the dependence of the thermal transport across the SQUIPT on the magnetic flux. 

Depending on the requirements, one can opt for low values of $T_S$ in order to increase the $OAR$ or choose high values of $T_S$ to maximize $G$. Since the behavior of the device is satisfactory both in terms of gain and output active range only in a limited range of supply temperatures, the use of materials with higher critical temperature for the ring of the SQUIPT could be beneficial in terms of device performances. Higher values of $T_{C-S_1}$ would guarantee wider $OAR$ and larger $G$ at $T_{IN}=Sens$. The maximum value of the gain in the active region at the optimal constant ratio $T_{C-S_1}/T_S\approx5.2$ rises linearly with the critical temperature of the SQUIPT for every value of $Sens$, as depicted in fig. \ref{Figure2}-d. Therefore, the PTA could potentially be used both at higher values of $T_S$ and $T_{IN}$ ensuring large $G$ and wide $OAR$ too. 

\section{Figures of merit}
In full analogy with electronics, we define particular figures of merit for the temperature amplifier. First of all, in our system the input-to-output thermal impedance $Z_{IN-OUT}^{th}$ is infinite. This arises from the double thermal-to-electrical-to-thermal transduction which ensures perfect heat decoupling between the input load and the output signal. Thereby, no heat current flows directly from the input lead to the output electrode.

\begin{figure*}[ht]
  \includegraphics {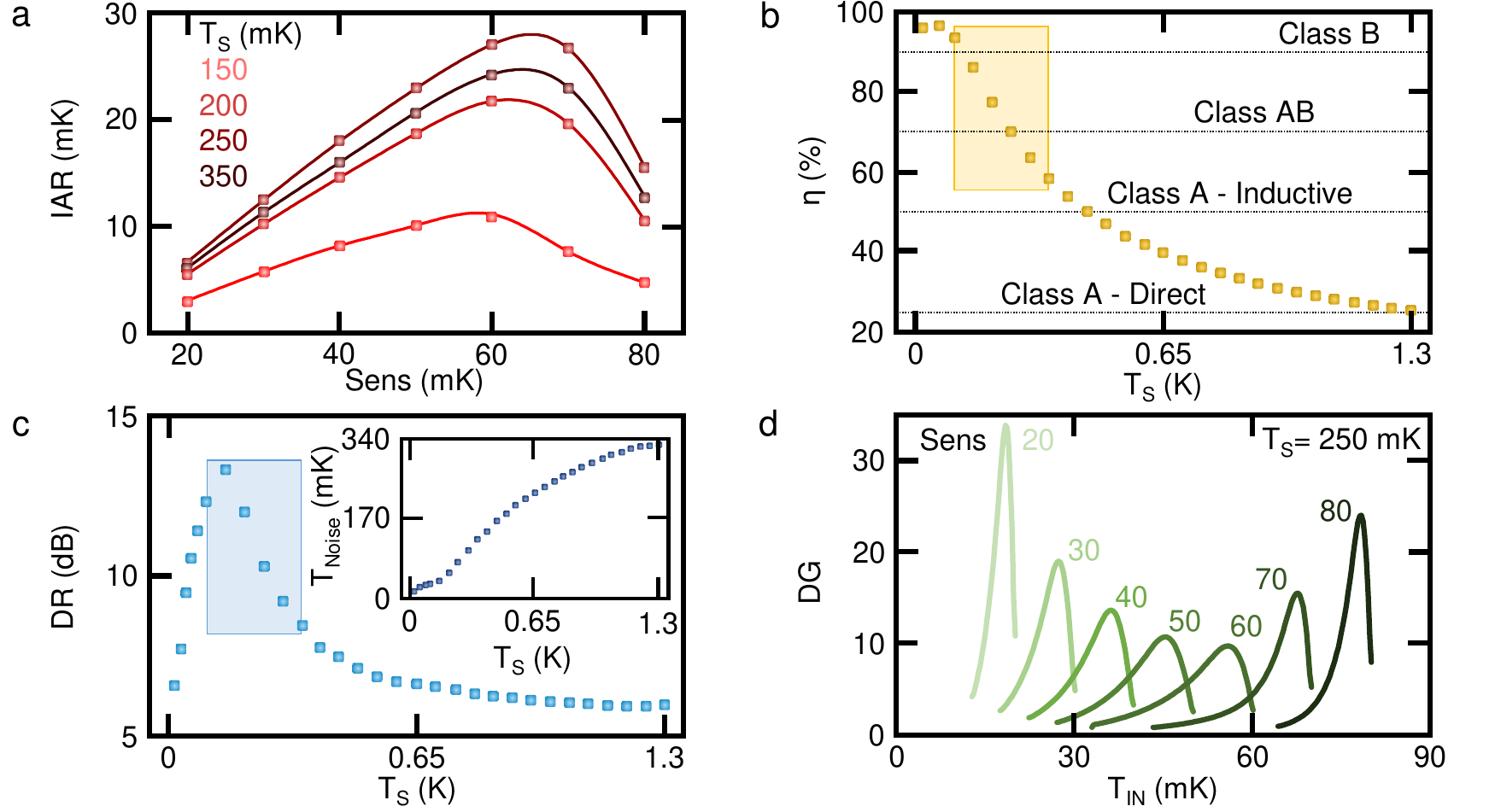}
  \caption{(a) Amplification range of the input temperature $IAR$ as a function of $Sens$ calculated for different values of $T_S$. (b) Efficiency $\eta$ as a function of $T_S$. The yellow rectangle depicts the area of maximum performances in terms of $G$ and $OAR$. Typical efficiencies of common voltage amplifiers are shown for a comparison. (c) Output dynamic range $DR$ as a function of $T_S$. The turquoise rectangle represents the area of maximum performances in terms of $G$ and $OAR$. Inset: output noise $T_{Noise}$ as a function of $T_S$. (d) Differential gain $DG$ as a function of $T_{IN}$ calculated at $T_S=250~$mK for different values of $Sens$.}
  \label{Figure3}
\end{figure*}

Another important parameter is the input amplification range that represents the interval of the input signal for which the output resides in the $OAR$. The length of this interval $IAR$ is defined as:
{\footnotesize
\begin{equation}
IAR=Sens-T_{IN}(T_{OUT_{MIN}}),
\label{eq:DeltaTin}
\end{equation}
}
where $T_{IN}(T_{OUT_{MIN}})$ is the value of the input temperature corresponding to the minimum value of the $OAR$. The $IAR$ is a function both of the $T_S$ and of $Sens$, as illustrated in fig. \ref{Figure3}-a. For small values of $T_S$ the $OAR$ is small and, hence, the $IAR$ is not extended too. By rising the supply temperature the $IAR$ enlarges till $T_S$ reaches about $250~$mK. A further increase of the supply temperature yields a softening of $\Delta_{S_1}(T)$, and a consequent compression of $OAR$, as already elucidated above. The reduction of the $OAR$ is mirrored in a narrowing of the $IAR$. The non-monotonic behavior of the $IAR$ with the $Sens$ comes from the competition between the two terms on the right side of Eq. (1) and can be ascribed to the thermoelectric element. One the one hand, the increase of $Sens$ naturally enlarges the $IAR$ by widening the total input temperature range. On the other hand, $I_T$ rapidly rises with $T_{IN}$, as illustrated in fig. \ref{Figure1}-c. The resulting magnetic flux $\Phi$ is modulated only for values of the input temperature approaching $Sens$, because $M$ is small and for the thermocurrents typical of narrow temperature gradients the flux always tends to zero. The latter effect manifests itself in lowering $IAR$ for increasing $Sens$ (see fig. \ref{Figure3}-a).

In our amplifier, the temperature is the potential used in the amplification. Hence we can define the efficiency $\eta$ as:
{\footnotesize
\begin{equation}
\eta=\frac{T_{OUT_{MAX}}}{T_S}\times 100,
\label{eq:Eta}
\end{equation}
}
where $T_{OUT_{MAX}}=T_{OUT}(T_{IN}=Sens)$. The efficiency reaches $\sim95\%$ for very small supply temperatures and monotonically decreases with rising $T_S$, as plotted in fig. \ref{Figure3}-b. The drop of $\eta$ can be explained with the closure of $\Delta_{S_1}$ and the growth of the losses through the phonons resulting from the temperature increase \cite{Giazotto2006}. In the region of best performances in terms of $OAR$, $G$ and $IAR$ (represented with the yellow rectangle in fig. \ref{Figure3}-b) the efficiency ranges from $\sim90\%$ to $\sim60\%$. These large $\eta$ values are comparable to analogous commercial electronic amplifiers. 

The $OAR$ provides a first and reliable estimate of the useful interval of the output signal. A more complete analysis employs the output dynamic range $DR$ defined as:
{\footnotesize
\begin{equation}
DR=20\times \log\left( \frac{T_{OUT_{MAX}}+T_{Noise}}{T_{Noise}}\right).
\label{eq:DR}
\end{equation}
}
The $DR$ widens by increasing supply temperature up to $T_S=150~$mK, because $T_{OUT_{MAX}}$ rises while $T_{Noise}$ is almost unaffected (as shown in the inset of fig. \ref{Figure3}-c). Further increase of $T_S$ enlarges  the noise with a steeper rate, while $T_{OUT_{MAX}}$ tends to level to a constant value. As a consequence, $DR$ decreases for values of $T_S$ approaching the SQUIPT critical temperature. Despite that, the PTA reaches the maximum performances in terms of $DR$ in the optimal region in terms of the other figures of merit, as depicted by the turquoise rectangle in fig. \ref{Figure3}-c.

Finally, we consider the differential gain, defined as:
{\footnotesize
\begin{equation}
DG=\frac{\mathrm d T_{OUT}}{\mathrm d T_{IN}}.
\label{eq:dG}
\end{equation}
}
At a fixed sensitivity, $DG$ displays a bell-like shape, as shown in fig. \ref{Figure3}-d. The height, width and position of the peak are sensitivity-dependent. In particular, for small and large values of $Sens$ the peak is high and narrow, while for intermediate sensitivities the peak is low and broad in $T_{IN}$. Since $DG$ is always greater than zero, the output signal is always a monotonically growing function of the input, as required for an amplifier.
\section{Conclusions}
We have proposed the phase-tunable temperature amplifier, which is the caloritronic counterpart of the voltage amplifier in electronics. The pivotal architecture proposed in this work constitutes the first fully-thermal on-chip device in coherent caloritronics, because the magnetic field necessary to control the thermal nano-valve (SQUIPT) is self-generated by the use of a thermoelectric element ($N-FI-S$ junction). The operating principle and the performances have been studied in detail paying specific attention to the experimental feasibility of geometry and material composition. The predicted input-to-output temperature conversion provides a maximum gain $G\approx11$ at small input signals which is mainly limited by the superconducting critical temperature $T_{C-S_1}$ of the Al-based nano-valve. In addition, we defined several figures of merit in full analogy with voltage amplifiers obtaining remarkable results especially in terms of output dynamic range $DR$ and efficiency $\eta$.
\acknowledgments
The authors acknowledge the European Research Council under the European Unions Seventh Framework Programme (FP7/2007-2013)/ERC Grant No. 615187 - COMANCHE and the MIUR under the FIRB2013 Grant No. RBFR1379UX - Coca for partial financial support. The work of E.S. is funded by a Marie Curie Individual Fellowship (MSCA-IFEF-ST No. 660532-SuperMag).
\section{Appendices} 
\subsection{N-FI-S junction}
The thermoelectric is a tunnel junction made of a normal metal $N$ at temperature $T_{IN}$, a ferromagnetic insulator $FI$ and a superconductor $S$ at $T_{BATH}$. The $FI$ layer operates a double action: it behaves as a spin filter with polarization $P=(G_{\uparrow}-G_{\downarrow})/(G_{\uparrow}+G_{\downarrow})$ where $G_{\uparrow}$ and $G_{\downarrow}$ are the spin up and spin down conductances \cite{Moodera2007}, and it causes the spin-splitting of the DOS of the superconductor by the interaction of its localized magnetic moments with the conducting quasiparticles in $S$ through an exchange field $h_{ex}$. Since the exchange interaction in a superconductor decays over the coherence length $\xi_0$ \cite{Tokuyasu1988}, we assume $S$ thinner than $\xi_0$ and a spatially homogeneous spin-splitted DOS \cite{Giaz2008}:
{\footnotesize
\begin{equation}\tag{A1}
N_{\uparrow,\downarrow}(E)=\frac{1}{2} \left| \Re\left\lbrack \frac{E+i\Gamma \pm h_{ex}}{\sqrt{(E+i\Gamma \pm h_{ex})^2-\Delta^2}} \right\rbrack \right|,
\label{eq:DOSsplit}
\end{equation}
}
where $E$ is the energy, $\Gamma$ is the Dynes parameter accounting for broadening \cite{Dynes1984}, and $\Delta(T_{BATH}, h_{ex})$ is the temperature and exchange field-dependent superconducting energy gap. The pairing potential is calculated self-consistently from the BCS equation \cite{Tinkham, Giaz2008}:
{\footnotesize
\begin{equation}\tag{A2}
\begin{split}
\ln \left(\frac{\Delta_0}{\Delta} \right)=\int_0^{\hbar\omega_D}\frac{f_+(E)+f_-(E)}{\sqrt{E^2+\Delta^2}}dE
\end{split}
\label{eq:GapEq}
\end{equation}
}
where $f_{\pm}(E)= \left \{ 1+\exp\frac{\sqrt{E^2+\Delta^2}\mp h_{ex}}{k_BT_{BATH}} \right \}^{-1}$ is the Fermi distribution of the electrons, $\omega_D$ is the Debye frequency of the superconductor, $\Delta_0$ is the zero-field and zero-temperature superconducting gap, and $k_B$ is the Boltzmann constant. The tunnel thermocurrent in the closed circuit configuration is only due to the temperature gradient and takes the form:
{\footnotesize
\begin{equation}\tag{A3}
\begin{split}
&I_T=\frac{1}{eR_T}\int_{-\infty }^{\infty }\left\lbrack N_+(E)+PN_-(E)\right\rbrack \\
&\left\lbrack f_N(E,T_{IN})-f_S(E,T_{BATH}) \right\rbrack dE,
\end{split}
\label{eq:It}
\end{equation}
}
where $e$ is the electron charge, $R_T$ is the tunnel resistance in the normal state, $N_{\pm}(E)=N_{\uparrow}(E)\pm N_{\downarrow}(E) $, $f_{N}(E,T_{IN})= \left \lbrack 1+\exp\left( E/k_BT_{IN} \right)\right \rbrack^{-1}$ and $f_{S}(E,T_{BATH})= \left \lbrack 1+\exp\left( E/k_BT_{BATH} \right)\right \rbrack^{-1}$ are the metal and superconductor Fermi functions, respectively.

\subsection{Temperature-biased SQUIPT}
We model the SQUIPT as a superconducting ring $S_1$ interrupted by a one-dimensional normal metal wire $N_1$ ($l\gg w,t$ where $l$, $w$ and $t$ are the wire length, width and thickness, respectively). The superconducting properties acquired by the wire through the proximity effect \cite{Holm1932} has been shown to be modulated by the magnetic flux $\Phi$ threading the ring \cite{Giazotto2010, Meschke2011}. Similarly, it has been recently shown that by interrupting the $S_1$-loop of the SQUIPT with a superconducting wire $S_2$, the superconducting properties of the latter are tuned by the magnetic flux threading the loop \cite{Virtanen2016}. Here we consider a hybrid superconductor-normal metal SQUIPT. Finally, a normal metal $N_2$ probe is tunnel-coupled to the wire through a thin insulating layer, and acts as output electrode. The DOS of the wire $N_{wire}$ is the real part of the quasi-classical retarded Green's function $g^R$ \cite{Rammer1986} obtained by solving the one-dimensional Usadel equation \cite{Usadel1970}. In the short junction limit (i.e. when $E_{Th}=\hbar D/l^2\gg \Delta_{0_{S_1}}$, where $E_{Th}$ is the Thouless energy, $\hbar$ is the reduced Planck constant and $D$  is the wire diffusion coefficient) the proximity effect is maximized, and the DOS is expressed by \cite{Strambini2014, Giazotto2010}:
{\footnotesize
\begin{equation}\tag{A4}
\begin{split}
N_{wire}(E,\Phi)=\left| \Re\left\lbrack \frac{E-iE_{Th}\gamma g_s}{\sqrt{(E-iE_{Th}\gamma g_s)^2+\left\lbrack E_{Th}\gamma f_s \cos \left( \frac{\pi \Phi}{\Phi_0} \right) \right\rbrack^2}} \right\rbrack \right|.
\end{split}
\label{eq:DOSwire}
\end{equation}
}
Above, $\gamma=R_{N_1}/R_{S_1N_1}$ is the transmissivity of the $S_1N_1$ contact (with $R_{N_1}$ resistance of the normal wire and $R_{S_1N_1}$ resistance of the $S_1N_1$ interface), $g_S(E)=\frac{E+i\Gamma_{S_1}}{\sqrt{(E+i\Gamma_{S_1})^2-\Delta_{S_1}^2}}$ is the coefficient of the phase-independent part of the DOS (with $\Gamma_{S_1}$ Dynes broadening parameter \cite{Dynes1984} and $\Delta_{S_1}$ BCS temperature dependent energy gap \cite{Tinkham}), $f_S(E)=\frac{\Delta_{S_1}}{\sqrt{(E+i\Gamma_{S_1})^2-\Delta_{S_1}^2}}$ is the coefficient of the phase-dependent part of the DOS, and $\Phi_0=2.068\times10^{-15}~$Wb is the magnetic flux quantum. The heat current $J$ tunneling from the $S_1N_1$ ring to the $N_2$ probe has been theoretically \cite{Giazotto2006, Strambini2014} and experimentally \cite{Giazotto2012, Mart2014, Fornieri2015} shown to depend on the temperatures of the ring $T_S$ and of the normal electrode $T_{OUT}$ through:
{\footnotesize
\begin{equation}\tag{A5}
\begin{split}
&J(T_S,T_{OUT},\Phi)= \\
&\frac{2}{e^2R_{T_1}}\int_{0}^{\infty}N_{wire}(E) \left\lbrack f_0(E,T_S)-f_0(E, T_{OUT}) \right\rbrack E dE
\end{split}
\label{eq:Jsquipt}
\end{equation}
}
where $f_0(E, T)=\left\lbrack1+\exp (E/k_BT)\right\rbrack^{-1}$ is the Fermi distribution of the quasiparticles in the ring for $T=T_S$ and in the probe for $T=T_{OUT}$.  The steady-state temperature of the probe $T_{OUT}$ depends on the thermal current flowing from $S_1N_1$ to $N_2$ and on the exchange mechanism occurring in $N_2$. Below $\sim1~$K the relaxation is mainly due to electron-phonon coupling \cite{Giazotto2006} and can be quantified as $J_{e-ph, N_2}(T_{OUT}, T_{BATH})=\Sigma~ V \left(T_{OUT}^n-T_{BATH}^n\right)$, where $\Sigma$ is the electron-phonon coupling constant, $V$ is the volume of the probe and the exponent $n$ depends on the disorder of the system. For metals, in the clean limit $n=5$, while in the dirty limit $n=4,6$ \cite{Giazotto2006, Strambini2014}. At the steady state by setting a constant temperature of the superconducting ring $T_S$ the output temperature of the nano-valve $T_{OUT}$ can be obtained by solving the following balance equation:
{\footnotesize
\begin{equation}\tag{A6}
\begin{split}
-J(T_S, T_{OUT},\Phi)+J_{e-ph, N_2}(T_{OUT}, T_{BATH})=0.
\end{split}
\label{eq:BalEq}
\end{equation}
}
\subsection{Materials and geometry}
The thermoelectric element is composed of $15~$nm of Cu as $N$, $1~$nm of EuS as $FI$ and $3~$nm of Al as $S$. Within this geometry the Al layer has typically: $T_C\approx3~$K, $\Delta_0\approx 456~\mu$eV and $\Gamma=1\times10^{-4}\Delta_0$. We consider an EuS layer characterized by: $P=0.95$, $h_{ex}=0.45~\Delta_0$ and $R_T=0.1~\Omega$. The superconducting coil originating the magnetic flux is made of $10~$nm thick aluminum and it is embedded in $10~$nm of Al$_2$O$_3$. The SQUIPT is made of a copper $N_1$ wire ($l=100~$nm, $w=30~$nm, $t=30~$nm) of diffusivity $D=1\times10^{-2}~$m$^2/$s, and of a $150~$ nm thick Al $S_1$ ring of radius $r_{SQUIPT}=5~\mu$m with $\Delta_{0-S_1}=200~\mu$eV, $T_{C-S_1}\approx1.32~$K and $\Gamma_{S_1}=1\times10^{-4}\Delta_{0-S_1}$. The transmissivity of the $S_1N_1$ contact is $\gamma=33$. The AlMn probe is tunnel-coupled to the proximized wire through a $1~$nm thick aluminum oxide layer ($R_{T_1}=100~$k$\Omega$). The parameters of the AlMn electrode are: $\Sigma=4\times10^{9}~$WK$^{-6}$m$^{-3}$ \cite{Fornieri2015}, $V=1\times10^{-20}~$m$^3$ and $n=6$ \cite{Giazotto2006, Fornieri2015}.

\end{document}